\documentclass[usenatbib]{mn2e}
\usepackage{natbib}
\usepackage{amsmath}
\usepackage{amssymb}
\usepackage{graphicx}
\usepackage{color}
\usepackage[colorlinks=true,linkcolor=darkblue,citecolor=darkblue,urlcolor=darkblue]{hyperref}
\newlength{\mystretch}
\definecolor{darkblue}{rgb}{0.05,0.0,0.5}

\newcommand{\Cincludegraphics}[1]{
\begin{center}
\includegraphics[width=0.45\textwidth]{#1}
\end{center}
}

\begin{document}
\title[Cosmology when living near the Great Attractor]{Cosmology when living near the Great Attractor}
\author[Wessel Valkenburg and Ole Eggers Bj{\ae}lde]{Wessel Valkenburg$^1$\thanks{w.valkenburg@thphys.uni-heidelberg.de} and Ole Eggers Bj{\ae}lde$^2$\thanks{oeb@phys.au.dk}\\
$^1$Institut f\"ur Theoretische Physik, Universit\"at Heidelberg \\ Philosophenweg 16, D-69120 Heidelberg, Germany\\
$^2$Department of Physics and Astronomy, Aarhus University \\ Ny Munkegade 120, DK-8000 Aarhus C, Denmark}

\maketitle
\begin{abstract}
If we live in the vicinity of the hypothesized Great Attractor, the age of the universe as inferred from the local expansion rate can be off by three percent.
We study the effect that living inside or near a massive overdensity has on cosmological parameters induced from observations of supernovae, the Hubble parameter and the Cosmic Microwave Background. We compare the results to those for an observer in a perfectly homogeneous $\Lambda$CDM universe. We  find that for instance the inferred value for the global Hubble parameter changes by around three per cent if we happen to live inside a massive overdensity such as the hypothesized Great Attractor. Taking into account the effect of such structures on our perception of the universe makes cosmology perhaps less precise, but more accurate.
\end{abstract}
\section{Introduction}

According to the standard model of cosmology, the universe is best described as being homogeneous and isotropic with an energy budget dominated by roughly 1/4 matter and 3/4 dark energy. The homogeneity and isotropy apply only on large scales, which in this context means roughly 100 Mpc. On much smaller scales the universe is obviously not homogeneous and isotropic because of local deviations from the global distribution of matter and possibly dark energy. The large scale homogeneity and isotropy of the universe have been demonstrated by several different observations, such as the Cosmic Microwave Background and the distribution of matter, the Large Scale Structure. At the same time, however, these observations are not as constraining when it comes to specifying our own position in the universe. If for instance we happened to live in a large local inhomogeneity such as a massive overdensity or underdensity this could lead to a potential bias in the inference of the global properties of the universe. Living in a region that is far from the average density, can strongly alter our perceived expansion history of the universe, as exemplified in the Large Local Void scenario~\citep{1995ApJ...453...17M,Tomita:1999qn,Alnes:2005rw,Enqvist:2006cg,GarciaBellido:2008nz,February:2009pv,Biswas:2010xm,Moss:2010jx} and in other studies combining smaller voids and Dark Energy in one model~\citep{Sinclair:2010sb,Marra:2010pg,Valkenburg:2011ty}.

One massive overdensity is the hypothesized Great Attractor \citep{LyndenBell:1988qs}. As the name suggests, the Great Attractor is a large extended mass distribution, which dominates the dynamics of the local universe by pulling in galaxies at large distances \citep{Erdogdu:2006nd,Haugboelle:2006uc}. It has been proposed to explain, among other things, the measured peculiar motion of the Milky Way galaxy, but its position and size remain largely unknown. In the literature the mass of the Great Attractor has previously been estimated to be somewhere between $M_{\rm GA}\,\sim\,10^{15}-10^{17} M_{\bigodot}$ and its redshift distance has been estimated to be somewhere between $cz_{\rm GA}\,\sim\,3000-5000$ km/s, indicating that we could be living in the outskirts of a supermassive cluster of galaxies with the Great Attractor in its center \citep{Kraan-Korteweg:1996,Kocevski:2005pj,Kocevski:2005kr}.

Several authors have previously looked for anisotropy in for instance supernova data \citep{Davis:2010jq,Colin:2010ds,Dai:2011xm} and have concluded that at low redshift peculiar velocities are of great importance and need to be accounted for. The reason for these peculiar velocities or bulk flows are assumed to be massive overdensities such as the Great Attractor or perhaps the Shapley Concentration at even further distance $cz_{\rm SC}\,\sim\,13000$ km/s.

In this paper we investigate the effect of living in a massive overdensity on the inferred global properties of the universe. The range over which such an overdensity extends in redshift space is too small to be of relevance for estimates of the equation of state of Dark Energy, as opposed to analyses in for example~\citep{deLavallaz:2011tj,Valkenburg:2011ty,Marra:2012}. It could affect estimates of the equation of state of dark energy through our peculiar velocity, which changes redshift relations and thereby propagates in all cosmological parameters, but this effect is found to be negligible~\citep{Davis:2010jq}. We show that for an observer living near the Great Attractor, mainly the locally observed expansion rate is observed to be different from the global expansion rate, and hence would lead to a wrong estimate of the age of the universe if the observer falsely assumes that he or she lives in an environment of  average density.

\section{Modelling the Great Attractor\label{sec:modga}}
\subsection{The metric}
As simplest approximation we describe the Great Attractor by a spherical collapse embedded in a $\Lambda$CDM universe, using the Lema\^itre-Tolman-Bondi (LTB) metric,
\begin{align}
ds^2=-dt^2+S^2(r,t)dr^2+R^2(r,t)(d\theta^2+\sin^2\theta\,d\phi^2),
\end{align}
where
\begin{align}
S(r,t)=&\frac{R'(r,t)}{\sqrt{1+2r^2k(r)\tilde M^2}},&
R(r,t)=&r\,a(r,t).
\end{align}
with the LTB equation,
\begin{align}
\left( \frac{\dot R}{R} \right)^2 =  \left( \frac{\dot a}{a} \right)^2 = & H^2(r,t)=\nonumber\\ \frac{8 \pi \tilde M^2}{3} & \left[   \frac{1}{a^3(r,t)} +  \frac{3k(r)}{4\pi a^2(r,t)}  + \frac{\Lambda}{8\pi\tilde M^2}    \right],\label{eq:LTBi}
\end{align}
where we chose a gauge for the radial coordinate,
\begin{align}
\int_0^r dr\,M'(r) \sqrt{1+2r^2k(r)\tilde M^2}=&4\pi\tilde M^2 r^3/3,\label{eq:gaugechoice}
\end{align}
such that the matter density is given by
\begin{align}
\rho(r,t)=&\frac{ \tilde M^2 r^2}{R'(r,t)R^2(r,t)}.\label{eq:rhodef}
\end{align}
This configuration is described by two functions: the curvature $k(r)$ and the Big-Bang time $t_{BB}(r)$. We choose $t_{BB}(r)\equiv0$.

We parametrize the curvature by
\begin{align}
k(r) =& k_{\rm max} W_3\left(\frac{r}{L},0\right) + k_b,
\end{align}
where $W_3(x,\alpha)$ is the third order of the function $W_n(x,\alpha)$ which interpolates from $1$ to $0$ in the interval $\alpha < x < 1$, while remaining $C^n$ everywhere, as defined in \cite{Valkenburg:2011tm}. Hence $k(r)$ is $C^3$ everywhere, such that the metric is $C^2$ and the Riemann curvature is $C^0$. The constant $k_b$ describes the background in which the spherical object is embedded. Since $k_b$ is a constant, the background is an exact Friedmann-Lema\^itre-Robertson-Walker solution, for all $r>L$. By construction, since $k(r)$ goes to a constant $k_b$ at finite radius, this curvature profile describes compensated structures: an overdensity surrounded by an underdense shell, matched exactly to the average background density at finite radius $r=L$.

In the following we write $r_*$ for an arbitrary $r>L$, simply to indicate a function's value in the FLRW background metric. Normalizing $a(r_*,t_0)=1$ where $t_0$ denotes today, we have
\begin{align}
k_b\equiv&\frac{4\pi}{3}\frac{\Omega_k(r_*)}{1-\Omega_k(r_*)-\Omega_\Lambda(r_*)}\label{eq:norm1}\\
\tilde M^2 \equiv& \frac{3 H^2(r_*,t_0) - {\Lambda}}{8\pi} \frac{1}{1+\frac{3k_b}{4\pi}}, \\
t_0\equiv& t(a(r_*,t_0)) ,
\end{align}
where $\Omega_\Lambda(r_*)\equiv \Lambda / 3H(r_*,t_0)^2$ and throughout the rest of this paper we take $\Omega_k(r_*)=k_b=0$.

We use the solution to Eq.~\eqref{eq:LTBi} described in \citep{Valkenburg:2011tm},
\begin{align}
t(a) - t_{BB}(r) =\nonumber\\%=&\frac{1}{\tilde M}\int_0^{a}\frac{\sqrt{\tilde a}\,d\tilde a}{\sqrt{ \frac{8 \pi}{3} +  2 k(r) {\tilde a}  + \frac{\Lambda}{3\tilde M^2}  {\tilde a^3 }}},\label{eq:t_LTB_full}\\
\frac{2}{\sqrt{3\Lambda}}\frac{(-1)^{-\frac{9}{2}}}{\sqrt{z_1 z_2 z_3}} & R_J\left(\frac{1}{a}-\frac{1}{z_1},\frac{1}{a}-\frac{1}{z_2},\frac{1}{a}-\frac{1}{z_3}, \frac{1}{a}  \right),\label{eq:t_LTB_full_soln}
\end{align}
where $R_J(x,y,z,p)$ is Carlson's Elliptic Integral of the Third Kind~\citep{1995NuAlg..10...13C}. The parameters $z_i$ are the three (complex) roots of $\frac{8\pi\tilde M^2}{\Lambda}+\frac{6 \tilde M^2 k(r)}{\Lambda} z_i + z_i^3=0$.

\subsection{Distance measures}
For a central observer, at $r=0$, owing to spherical symmetry all (null) geodesics going through that point are purely radial. Therefore, the physical surface spanned by a collection of such geodesics describing a physical infinitesimal light cone is always given by $dA=R^2(r,t) d\Omega$, where $d\Omega$ denotes the solid-angular separation. Hence, the angular diameter distance is equal to $d_A(z) = R(r(z),t(z))$.

For an off-center observer, the only symmetry left is the rotational symmetry along the axis that connects the observer and the center of the system. Photons along an actual cone with its tip at the observer and centered on the radial axis, no longer travel on radial geodesics, {\em i.e.} their angular coordinates are not constant along the geodesic. This implies that it is possible for lensing effects to occur along the light cone, altering the beam width and thereby the angular diameter distance to an object. Hence, one has to solve for the optical equation, which gives the beam width along a photon geodesic~\citep{Brouzakis:2006dj,Brouzakis:2007zi,Valkenburg:2009iw},
\begin{align}
	\frac{d\xi}{d\lambda}&= 4 \pi \tilde M^2 r^2 A \left(\frac{d\phi}{d\lambda}\right)^2
 		\left(\frac{1}{R'(r,t)} - \frac{1}{R(r,t)} \right)\label{eq:exactda1}\\
	\frac{d^2\sqrt{A}}{d\lambda^2}&=-\frac{1}{2}\sqrt{A}\,\,\rho(r,t) \left(\frac{dt}{d\lambda}\right)^2 - \frac{\xi^2}{A^{3/2}},\label{eq:exactda2}
\end{align}
with the initial conditions for integration from the observer back in time,
\begin{align}
	\left.\frac{d\sqrt{A}}{d\lambda}\right|_{\lambda = 0}&=\sqrt{\Omega_{\rm obs}},\\
	\left.\sqrt{A}\right|_{\lambda = 0}&= 0,\label{eq:inisqa}\\
	\left.\xi\right|_{\lambda = 0}&=0,
\end{align}
where $\lambda$ is the affine parameter along the geodesic and $\Omega_{\rm obs}$ is the solid angle under which the observer observes the beam.
Here, $\xi\equiv A\sigma$, with $\sigma$ the beam shear.
The geodesics themselves are then given by,
\begin{align}
\frac{dz}{d\lambda}=&\,-\frac{\dot R'(r,t)}{R'(r,t)}\left((z+1)^2-\frac{c_\phi^2}{R^2(r,t)}\right)-c_{\phi}^2\frac{\dot R(r,t)}{R^3(r,t)},   \label{eq:geo1}\\
\frac{dt}{d\lambda}=&\,z+1,\\
\frac{dr}{d\lambda}=&\,\frac{\sqrt{1+2r^2k(r)\tilde M^2}}{R'(r,t)}\sqrt{(z+1)^2-\frac{c_{\phi}^2}{R^2(r,t)}}, \\
\frac{d\phi}{d\lambda}=&\,\frac{c_\phi}{R^2(r,t)} ,\label{eq:geo2}\\
z(0)=&\,0, \,\,\,\, t(0)=\,t_0, \,\,\,\, r(0)=\,r_{\rm obs}, \,\,\,\, \phi(0)=\,\phi_{\rm obs},
\end{align}
where we confine the geodesics to a plane without loss of generality. The parameter $c_{\phi}$ is the only parameter describing angular motion~\citep{Marra:2007pm}, so we can confine the analysis to radial geodesics by setting $c_{\phi}=0$. Following this choice, we see from Eq.~\eqref{eq:exactda1} that for radial beams the shear vanishes, as one expects given the rotational symmetry along a radial geodesic.

One now finds $d_A(z)$ on the radial geodesic crossing the position of an off-center observer at $t_0$ by integrating the specified equations, to find $d_A(z(\lambda))=\sqrt{\frac{A(z(\lambda))}{\Omega_{\rm obs}}}$, which itself is independent on the initial choice for $\Omega_{\rm obs}$. It is in practice convenient to choose $\Omega_{\rm obs}=1$.

Since we have photon conservation along the geodesic, the luminosity distance is given by $d_{L}(z)\equiv (1+z)^2 d_A(z)$. The distance modulus for an observed supernova is then $\mu\equiv 5 \log_{10} \left[{d_L}/{10 \mbox{pc}}\right]$.

\section{Fitting to data}
\subsection{Data and parameter priors}
The purpose of this work is to give a proof of concept, explicating the possibility that an observer living amongst in-falling matter observes altered distance measures. We focus on radial geodesics, both outward and inward, with the latter crossing $r=0$. Even though the observed supernovae are spread throughout the sky, we will fit them as if they are all on one unique angular position on the sky. Similarly, we will assume that the distance to the CMB in all directions is given by the distance to the surface of last scattering on the one considered radial geodesic. This certainly is an oversimplification, but we leave a full sky computation for possible future work, focussing on a proof of concept for now.

We perform fits against simultaneously Supernovae, Hubble parameter observations and the Cosmic Microwave Background. We use supernovae from the Sloan Digital Sky Survey Supernovae data release~\citep{Kessler:2009ys}.

We use the Hubble parameter constraints from~\cite{Riess:2009pu}, where one fits an observed distance at a single redshift of $z=0.04$. In~\cite{Riess:2009pu} this led to the conclusion $H_0=74.2\pm3.6$, but as we will see different values can fit the constraints for different cosmologies.

For the CMB we use the WMAP 7-year data release~\citep{Komatsu:2010fb}. We fit the CMB with an effective FLRW observer for whom the physics at the surface of last scattering is identical to that of the observer in the LTB metric, but with a corrected observation time different from $t_0$, such that the angular diameter distance to the surface of last scattering corresponds to that for the LTB observer~\citep{Biswas:2010xm,Moss:2010jx}. We use the full spectrum, although this means that the theoretical spectrum carries potentially wrong secondary anisotropies, such as the late-time integrated Sachs-Wolfe effect and CMB lensing.

Effectively this means that we have angular diameter distances measured at $z=0.04$ (HST), $0.024 < z < 1.55$ (SN) and $z=1089$ (CMB). Note that as a consequence $H_{0, {\rm out}}\equiv H(r_*,t_0)$ is not directly related to the observed value from the HST, which is given by roughly \mbox{$d_A(z=0.04)/0.04$}. The parameter $H_{0,{\rm out}}$ describes the expansion rate outside the LTB metric, and is relevant for the observer inside the LTB metric only for defining the age of the universe.

We use {\sc cosmomc}~\citep{Lewis:2002ah} for the Monte-Carlo sampling, {\sc VoidDistancesII}~\citep{Biswas:2010xm,Marra:2012} for the geodesics integration and distance calculations, {\sc CAMB}~\citep{Lewis:1999bs} for the calculation of the {\em effective} CMB spectrum, and {\sc ColLambda}~\citep{Valkenburg:2011tm} for the exact $\Lambda$LTB solution of the metric of the collapsing object embedded in $\Lambda$CDM.

Since we fit against HST, SN, and CMB, we allow $\Omega_{DM} h^2$, $\Omega_{b} h^2$, $H_{0, {\rm out}}$ plus the primordial spectral parameters $\log A_S$ (scalar amplitude), $n_S$ (scalar tilt) and $\alpha_S$ (running of the scalar tilt) and the optical depth to reionization $\tau$ to vary, describing the background with $\Omega_k\equiv0$ and $\Omega_\Lambda\equiv1-\Omega_m$. We use as a pivot scale for the primordial power spectrum $k_{\rm pivot} = 0.002$ Mpc$^{-1}$. Next to these standard parameters, we allow $r_{\rm proper, GA}$, $(\rho_{\rm in} - \rho_{\rm out}) / \rho_{\rm in}$ and $r_{\rm proper, obs}$ to vary, describing the collapsing structure and the position of the observer (see below). Note the denominator in $\frac{\rho_{\rm in} - \rho_{\rm out}}{  \rho_{\rm in}}$, which makes this quantity strictly smaller than unity, approaching unity for an infinite overdensity. We make this choice such that this parameter has a finite prior volume and does not bias the Monte-Carlo Markov Chains toward senseless results. The three last mentioned  parameters are defined as follows:
\begin{align}
r_{\rm proper, obs}& \equiv  \int_0^{r_{\rm obs}} S(r,t_0) dr ,\label{eq:robs}\\
r_{\rm proper, GA} &\equiv  \int_0^{r_{\rm GA}} S(r,t_0) dr , \label{eq:rga} \\
\frac{\rho_{\rm in} - \rho_{\rm out}}{  \rho_{\rm in}}&\equiv \frac{\rho(r=0,t_0)-\rho(r_*,t_0)}{\rho(r=0,t_0)},\label{eq:delta_inout}
\end{align}
where $r_{\rm obs}$ is the coordinate radius at which the observer resides, and $r_{\rm GA}$ is the coordinate radius at which $\rho(r,t_0)-\rho(r_*,t_0)$ changes sign, that is, the radius where the density transits from overdense to underdense. This radius always exists for the present configuration, because of the compensating underdense shell surrounding the Great Attractor, owing to which the exact matching to the FLRW metric occurs at a finite radius $r=L$.
All conversions to go from the three parameters~(\ref{eq:robs}--\ref{eq:delta_inout}) to the set $\{ L,k_{\rm max},  r_{\rm obs}\}$, which we encounter later in the analysis, are performed numerically.
\begin{table}
\begin{center}
\begin{tabular}{c}
\begin{tabular*}{0.4\textwidth}{c}
\hline
Flat priors\\ \hline \hline
\begin{tabular}{rcl}
$0.005 <$&$\Omega_{\rm b}h^2$ & $<0.1$ \\
$0.01 <$&$\Omega_{\rm dm}h^2$ & $<0.99$ \\
$0.2 <$ &$\frac{H_0}{\mbox{\tiny 100 km s$^{-1}$~Mpc$^{-1}$}}$ & $<1.2$\\
$0.01 <$ & $\tau$ & $ < 0.8$ \\
$2.7 <$ & $\log 10^{10} A_S$ & $ < 4$ \\
$0.5 <$ & $n_S$ & $ < 1.5$ \\
$-0.2 <$ & $\alpha_S$ & $ < 0.2$ \\
$0$  $<$&$r_{\rm obs} /L$ & $<1$\\
$0$~Mpc $<$&$r_{\rm GA} $ & $<500$~Mpc\\
$0 <$&$\frac{\rho_{\rm in} - \rho_{\rm out}}{  \rho_{\rm in}}$ & $<1$
\end{tabular}\\ \hline
\end{tabular*}\\ \\
\begin{tabular*}{0.4\textwidth}{c}
\hline
Additional constraints\\ \hline \hline
\begin{tabular}{rl}
$H(r=0,t_0)$&$<0$
\end{tabular}\\ \hline
\end{tabular*}
\end{tabular}
\end{center}
\caption{Priors imposed on the parameters in the numerical analysis. For the primordial power spectrum we use a pivot scale $k_{\rm pivot} = 0.002$ Mpc$^{-1}$. Note  the denominator $\frac{\rho_{\rm in} - \rho_{\rm out}}{  \rho_{\rm in}}$, which makes this quantity strictly less than one, and it approaches one for an infinite overdensity. }\label{tab:priors}
\end{table}
The priors on all the parameters are listed in Table~\ref{tab:priors}. Note that the constraint that $H(r=0,t_0)<0$ is actually very constraining on the central density $\frac{\rho_{\rm in} - \rho_{\rm out}}{\rho_{\rm in}}$, in practice meaning roughly $\frac{\rho_{\rm in} - \rho_{\rm out}}{  \rho_{\rm in}}>0.9$

\subsection{Results}
In the following we compare three MCMC runs that give parameter estimates: an observer living in (fitting the data with) an exact FLRW-universe ($\Lambda$CDM), an observer living in the outskirts of the Great Attractor at varying radius (fitting the data with the model described above), facing outward, and the same observer observing the universe {\em through} the center of the overdensity, {\em i.e.} facing in the opposing direction. In practice we integrate photon geodesics backward in time, starting from the observer. We let the coordinate $r$ grow backward in time, such that we obtain the universe for the inward facing observer by simply putting that observer at a negative radius $r$.

\begin{figure}
\Cincludegraphics{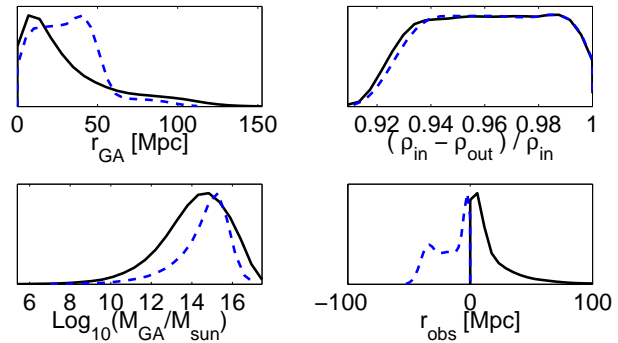}
\caption{Marginalized 1D posterior parameter likelihoods for the last three parameters in Table~\ref{tab:priors}, describing the Great Attractor and the position of the observer in it. Also shown is the likelihood function of the Great Attractor mass. The possibly negative $r_{\rm obs}$ reflects the inward facing observer (dashed, blue), as the geodesic integration uses a growing radial coordinate backward in time. }\label{fig:2}
\end{figure}

The last three parameters in Table~\ref{tab:priors} are only relevant for the case of the observer in the Great Attractor. Let us go through the constraints on these parameters first, to clarify the scenarios that we are considering. In Fig.~\ref{fig:2} we present the marginalized 1D posterior parameter likelihoods under the priors listed in Table~\ref{tab:priors} for those three parameters as well as the mass of the Great Attractor. We define the mass as,
\begin{align}
M_{\rm GA}\equiv 4\pi \int_0^{r_{\rm GA}} dr\, \sqrt{-g}\rho(r,t_0).
\end{align}
The lines in Fig.~\ref{fig:2}  correspond to the observer facing outward (solid, black) and the observer facing inward (dashed, blue). These posteriors show that in order for the perceived expansion history to be in agreement with observations, the overdensity can extend over a radius up to 100 Mpc, with the probability peaking around $20\sim50$ Mpc. The observer similarly lives at a distance varying from zero to 100 Mpc from the center, with the probability peaking around $10\sim20$ Mpc. The negative $r_{\rm obs}$ reflects the fact that the blue lines represent the observer that faces inward, through the center of the overdensity, because we take redshift $z$ to grow with growing radius $r$. The central density of the overdensity is mostly constrained by the condition that the overdensity is collapsing today, such that the only constraint is $0.9 \lesssim \frac{\rho_{\rm in} - \rho_{\rm out}}{  \rho_{\rm in}}<1$. Note that the density profile smoothly interpolates from overdense to underdense to the average density, such that the central density contrast is not representative for the average density contrast smoothed over the whole LTB patch. Apparently, the condition of collapse in the center implies that the over-density typically has a mass of around $10^{15} M_{\bigodot}$. Comparing these numbers with what we expect for the real Great Attractor, we first of all note that the extent of the the Great Attractor, our distance from its center as well as the mass are all unknown. However, an extent of roughly 50 Mpc matches perfectly with the expected size of some of the largest structures in the universe. We expect the Great Attractor to belong to this group of structures. Concerning the distance at which we live from the center of the Great Attractor, we would expect the distance to be somewhat greater than $10\sim20$ Mpc, but again the precise number is largely unknown. Finally a mass of order $10^{15} M_{\bigodot}$ matches well with the expected mass of the Great Attractor, as mentioned in the introduction. We conclude that the analysis is in agreement with the expected numbers for the real Great Attractor.

\begin{figure}
\Cincludegraphics{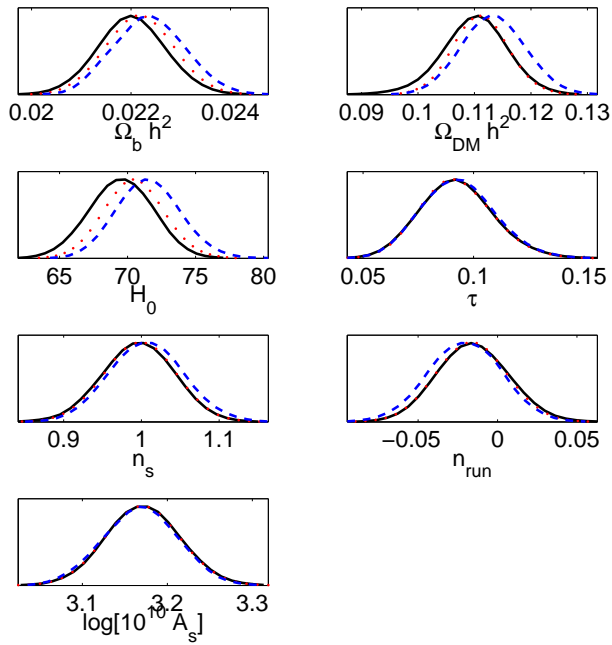}
\caption{Marginalized 1D posterior parameter likelihoods for the first seven parameters in Table~\ref{tab:priors}, describing the cosmology, for the observer facing outward from the Great Attractor (solid, black), facing inward (dashed, blue) and an observer in a pure $\Lambda$CDM universe (dotted, red). The parameter that is affected most significantly is $H_0$.}\label{fig:all}
\end{figure}

In Fig.~\ref{fig:all} we show the marginalized 1D posterior parameter likelihoods for the cosmological parameters under the priors listed in Table~\ref{tab:priors}. The lines correspond to $\Lambda$CDM (dotted, red), an observer living off-center of the Great Attractor, facing outward (solid, black), and a similar observer facing inward through the Great Attractor (dashed, blue). As is clearly visible, living near the Great Attractor introduces small shifts in all parameters compared to the $\Lambda$CDM observer. We focus on the change in the inferred value of $H_0=H_{0,{\rm out}}$ since this is the only parameter change which has a statistically interesting proportion. The observer in the Great Attractor lives in a universe with a shifted global value for $H_0$.

\begin{figure}
\Cincludegraphics{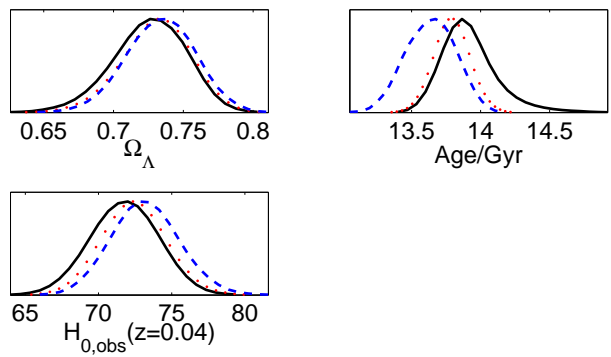}
\caption{Marginalized 1D posterior parameter likelihoods for derived parameters, describing the cosmology, for the observer facing outward from the Great Attractor (solid, black), facing inward (dashed, blue) and an observer in a pure $\Lambda$CDM universe (dotted, red). The change in $H_0$ (in Fig.~\ref{fig:all}) propagates by changing the values of $H_{0,{\rm obs}}$ and the age of the universe.}\label{fig:3}
\end{figure}

What this means can be seen in Fig.~\ref{fig:3}, where we show the derived parameters. For the outward facing observer in the Great Attractor the age of the universe is slightly higher. Should the observer not take into account that he is actually living inside a collapsing structure, he would infer the wrong value for the age of the universe. The change is of order of a few per cent. 

\begin{figure}
\begin{center}
\includegraphics[width=0.3\textwidth]{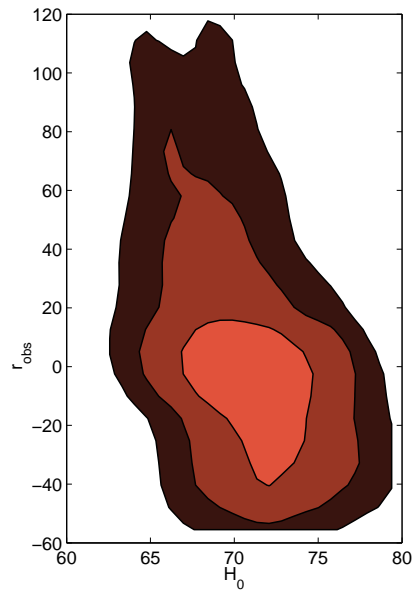}
\end{center}
\caption{Marginalized 2D posterior parameter likelihood contours for $H_0$ and $r_{\rm obs}$, where negative $r_{\rm obs}$ implies that the observer looks through the center, which explicitly shows the slight degeneracy between these two parameters. This information is obtained form a separate MCMC-run with the prior $-1<r_{\rm obs}/L<1$. The inner-most contour corresponds to the 68\% confidence level (c.l.), the second contour to 95\% c.l., and the outer-most contour to 99.7\% c.l.}\label{fig:contour}
\end{figure}

To show more explicitly the relation between $H_0$ and $r_{\rm obs}$, we performed a separate run with $-1<r_{\rm obs}/L<1$ and plot the 2D marginalized posterior probabilities for $H_0$ and $r_{\rm obs}$, including the 99.7\% confidence level (c.l.) contours, in Fig.~\ref{fig:contour}. This way we see the degeneracy between these two parameters that one finds if future  information on the Great Attractor based on astrophysical observations changes our prior on the size of the Great Attractor and our position amongst its infalling matter, for example fixing its size to something larger and our position amongst infalling matter at higher radii than what we find to be the best fit model at this moment. Including such information as a datapoint could push the good fit region into the outer contour, where one finds a strong impact on $H_0$ that would be missed if we focus only on the 1D marginalized likelihoods of Fig.~\ref{fig:all}

\subsection{Interpretation}
\begin{figure}
\Cincludegraphics{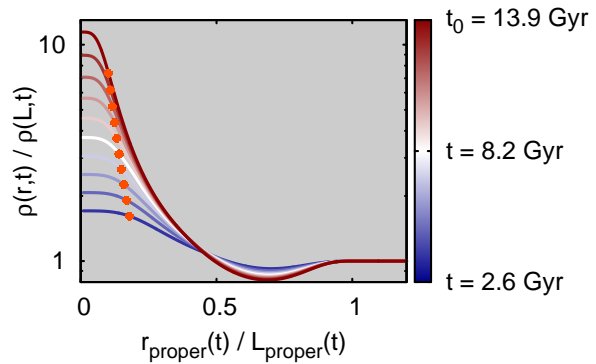}
\caption{An illustration of the relative matter density and of the position of the observer (the orange dot for illustration) as a function of time. Time increases according to the colour coding in the legend.}\label{fig:observer}
\end{figure}
\begin{figure}
\Cincludegraphics{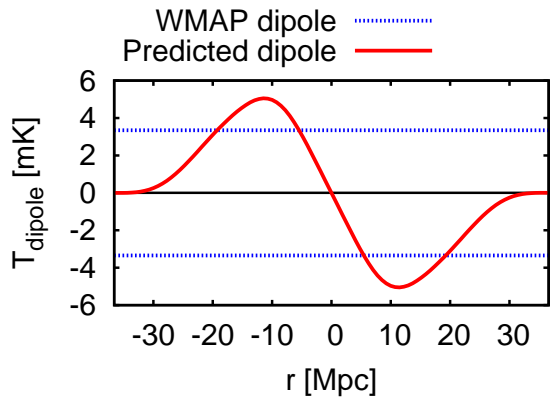}
\caption{The CMB dipole caused by the peculiar velocity of an observer living at various radii in a typical Great Attractor as modeled here. The predicted dipole is nowhere in violation of the observed dipole, and the magnitude is in perfect correspondence with observations. The sign of the dipole only reflects a direction, which is invisible when looking at multipole power.}\label{fig:dipole}
\end{figure}
To explain why the global value of $H_0$ changes when the observer lives in an overdensity, in Fig.~\ref{fig:observer} we illustrate the matter density as a function of time and proper distance (radius),
\begin{align}
r_{\rm proper}& \equiv\int_0^{r} S(r',t_0) dr' ,\\
L_{\rm proper}& \equiv\int_0^{L} S(r',t_0) dr' ,
\end{align}
and with an orange dot in the same figure we indicate the position of the observer, $r_{\rm obs}$ as defined in Eq.~\eqref{eq:robs}. That is, the observer sits at a constant coordinate radius, but physically has a motion with respect to the center of the Great Attractor; the coordinates are comoving with the dust. This picture clarifies how the collapsing structure plays a role in distance measures: the underdense sphere surrounding the Great Attractor expands faster than its surroundings, while the core collapses.
The observer hence looks at the universe through a lens.

Let us compare two observers, one facing through the underdense shell only, and one living in the same cosmology but without the inhomogeneity (the FLRW observer). A beam that converges at the observers, will for the former observer be relatively expanded by the underdensity. Therefore, if both observers see a beam with the same solid angle, for the former observer this will correspond to a smaller physical size at some fiducial redshift than for the latter observer: the underdense shell has increased the angular size of the beam such that it appears to come from a larger structure, which for the latter observer simply corresponds to an actually larger structure. Since  $d_{\rm A}=\sqrt{A/\Omega}$, for the former observer the angular diameter distance to a certain point will be {\em smaller} than for the latter observer. Since also $d_{\rm A}\propto H_0^{-1}$, the decrease in angular diameter distance can be compensated by lowering the overall expansion rate of the universe. And indeed, we see in Fig.~\ref{fig:all} that the observer facing outward, through the underdense shell, favours smaller values of $H_0$. Apparently for the observer that faces through the collapsing core, the effect is inverted. This can be understood by considering the total mass along the line of sight, which for this observer is above average, while for the outward facing observer it is below average. The average mass in this context is the mass along the line of sight the aforementioned FLRW observer sees. Hence, the inward facing observer favours higher values of $H_0$, as is indeed reflected in Fig.~\ref{fig:all}.

Naively, one could say that the integrated expansion rate along the line of sight is above average for the outward facing observer, and below average for the inward facing observer, where the FLRW observer defines the average. And since the angular diameter distance depends directly on the integrated expansion along the line of sight, and is strongly constrained by the CMB, this deviation in integrated expansion rate must be compensated by the global expansion rate of the universe.

Since the parameter constraints from the CMB are a tight interplay between the different parameters  (cf.  Table 3 in~\cite{Vonlanthen:2010cd}), some cosmological parameters have a shift in value, following the changed value of $H_0$, when we compare the inward and outward facing observers and the FLRW observer, in Figs.~\ref{fig:all}~and~\ref{fig:3}.

 Since the observer in this analysis is in-falling with the matter, one may wonder if the observers velocity is in agreement with current bounds on our velocity, which mainly come from the observed CMB dipole. By placing observers at different radii and comparing the redshift to a last scattering surface at fixed cosmic time $t_{\rm LSS}$, one can deduce the observed dipole for such an observer. In Fig.~\ref{fig:dipole} we show this dipole for an observer at different radii for a typical Great Attractor configuration. The peculiar velocity of the infalling observer predicts a CMB dipole which is in perfect agreement with observations.

\section{Discussion}
In this analysis we have investigated the effect of living inside a massive overdensity on the inferred global properties of the universe. We have demonstrated that, given the data used in the numerical analysis, an observer living inside a massive overdensity lives in a universe with slightly different global properties than an observer outside the overdensity in a smooth $\Lambda$CDM universe. The most interesting change is a slight change of the global value for the Hubble parameter $H_0$, which is roughly three per cent smaller if the observer lives inside a massive overdensity. This translates into a slightly higher age of the universe. All changes are small, but large enough to be taken into account in an era of precision cosmology, thus making the shift to `accuracy cosmology'.

There is a number of simplifications in this work, that could be addressed in future work in order to make the analysis more robust.

One simplification is the fact that we only consider radial geodesics in two directions: outward and crossing the center. Since these directions showed opposing effects on the cosmological parameters, one might guess that a full sky analysis, integrating geodesics in all directions, leads to a zero result. One must however remember that the observer is in all directions surrounded by in-falling matter, characterized by the underdense shell with an above-average expansion rate, such that even if the average effect cancels (as is the case in our present findings), there are most likely no directions in which the surrounding underdense shell has {\em no} effect on the distance measures. Moreover, the center of the Great Attractor has a small solid angle on the sky, such that in most directions the observer sees infalling matter, and it remains to be seen if the overall effect cancels or not.

A second simplification is the ignoring of effects of the Great Attractor on the CMB power spectrum. We only considered a simplified dipole. One could imagine that the quadrupole is too large to agree with observations, however this will strongly depend on the position of the observer: an observer close to the center really only sees a dipole~\citep{Alnes:2006pf}. Another signature that could show up in the CMB, even if the quadrupole were in agreement with observations,  is a correlation between different multipoles in the angular power spectrum of the CMB~\citep{Amendola:2010ty}, owing to the fact that the Great Attractor acts as one big lens.

\section*{Acknowledgements}
OEB thanks the Villum Foundation for support. WV acknowledges funding from DFG through the project TRR33 ``The Dark Universe''.

\bibliographystyle{apa}
\bibliography{refs}
\end{document}